# Satellite Navigation For The Age of Autonomy

Tyler G.R. Reid, Bryan Chan, Ashish Goel, Kazuma Gunning, Brian Manning,
Jerami Martin, Andrew Neish, Adrien Perkins, Paul Tarantino
*Xona Space Systems*
San Mateo, CA & Vancouver, BC
{tyler, bryan, ashish, kaz, brian, jerami, andrew, adrien, paul}@xonaspace.com

*Abstract*—Global Navigation Satellite Systems (GNSS) brought navigation to the masses. Coupled with smartphones, the blue dot in the palm of our hands has forever changed the way we interact with the world. Looking forward, cyber-physical systems such as self-driving cars and aerial mobility are pushing the limits of what localization technologies including GNSS can provide. This autonomous revolution requires a solution that supports safety-critical operation, centimeter positioning, and cyber-security for millions of users. To meet these demands, we propose a navigation service from Low Earth Orbiting (LEO) satellites which deliver precision in-part through faster motion, higher power signals for added robustness to interference, constellation autonomous integrity monitoring for integrity, and encryption / authentication for resistance to spoofing attacks. This paradigm is enabled by the 'New Space' movement, where highly capable satellites and components are now built on assembly lines and launch costs have decreased by more than tenfold. Such a ubiquitous positioning service enables a consistent and secure standard where trustworthy information can be validated and shared, extending the electronic horizon from sensor line of sight to an entire city. This enables the situational awareness needed for true safe operation to support autonomy at scale.

*Keywords—Autonomous Vehicles, Aerial Mobility, UAS, Low Earth Orbit (LEO), New Space, GNSS, Localization, Security;*

## I. INTRODUCTION

Satellite navigation has empowered our society, from the directionally challenged individual to the complex systems we rely upon. Global Navigation Satellite Systems (GNSS) find utility in nearly all facets of modern life, from the source of time for our communications networks to the source of safety-critical positioning in civil aviation. The now more than five billion [1] GNSS-enabled smartphones in our pockets have removed the mystery from getting from point A to point B, forever changing the way we interact with our physical world. Another revolution is on the horizon, one which promises disruption of transportation, mobility, and safety. Autonomous systems in the form of self-driving cars, autonomous aerial platforms, mobile robotics, and others are on the rise, targeting improved access to the mobility of people, goods, and services. This transformation is one whose complexity demands more than what navigation systems today can provide. Here, we present a vision for a connected and autonomous future which leverages investment in new space infrastructure to create a system for a unified and ubiquitous backbone for navigation services that are robust, reliable, and secure.

To understand, where navigation is headed, we first turn to the past and begin with an assessment of the historical trend. This includes contemporary drivers of new needs in accuracy, coverage, and capability as well as the technologies developed in the evolution from the sextant to the now more than one hundred navigation satellites in service today. This shows a clear trend: In the last century, there has been an order of magnitude improvement in location accuracy every thirty years. Each step has required investment in new infrastructure to reach new capabilities. With meter-level positioning first widely available in the mid-1990s with GPS, this implies that the mid 2020s will demand decimeter, or better, performance.

Autonomous systems are one of many coming applications that drive this need, where it is estimated that 10 cm, 95% accuracy in position will be required for self-driving cars [2]. Several technologies are emerging to meet this challenge. LiDAR, computer vision, radar, and GNSS are all striving towards this requirement. Though some have shown progress in meeting these needs in certain circumstances or conditions, they all struggle to fully solve the problem to the level of reliability, safety, and security that is needed. LiDAR and vision struggle in inclement weather due to absorption or scattering and is further hindered by occlusions. Radar, though more impervious to weather, is limited by sensor noise and resolution. LiDAR, vision, and radar approaches also require a data intensive localization map layer which must be maintained and updated frequently and has been identified as a major risk of the technology.

GNSS approaches have also seen substantial investment in the Advanced Driver Assistance System (ADAS) and full autonomous driving domains [3]. There is now continent-scale deployment of GNSS monitoring stations which target widespread accuracy at scale, delivering correction services via cellular connectivity. Though accuracy is approaching the needs of autonomy, other risks remain with interference and cybersecurity. Radio Frequency (RF) interference is a growing threat on the road. Typically motivated by privacy concerns, low-cost GNSS jammers are a popular means of disrupting fleet tracking. More than 50,000 disruptions were recorded in Europe alone between 2016 – 2018 [4]. The cybersecurity of GNSS is another emerging threat. Civil GNSS signals are unencrypted and unauthenticated, leaving vulnerabilities that can be exploited with counterfeit (spoofed) signals. In 2019, this vulnerability was demonstrated in autonomy with a staged spoofing attack on a Tesla Model S and Model 3, creating unsafe behavior of the autopilot [5].

The challenge facing the auto makers is how to achieve localization requirements on accuracy, availability, integrity, continuity, scalability, and security while maintaining reasonable Cost, Size, Weight, and Power (CSWaP). Furthermore, these systems will interoperate in our cities, necessitating a common standard. The safest maneuvers are informed with the most complete picture of the surroundings. This requires going beyond the line of sight of vehicle sensors and creating situational awareness at city levels. This demands an environment of collaborative data sharing through broadband connectivity and vehicle communication. This allows vehicles and infrastructure to act collectively, improving safety and reducing the risk of collision. Such data sharing is only effective if there is an agreed upon standard and datum with appropriate measures for data security.

Investment in infrastructure for the establishment of such a standardized system offers the potential for the most economical and lightweight navigation solution for the end user. Satellite navigation offers a ubiquitous reach with established global datums and seems to be the logical choice for such a universal standard. Though navigation accuracy is nearing the requirements for autonomous driving with GNSS through subscription-based correction services and capable receiver chipsets, autonomy has elevated expectations on navigation services with respect to resilience in the face of interference and cybersecurity.

To meet these demands, we propose a navigation service from Low Earth Orbiting (LEO) small satellites. As shown in Fig. 1, compared to GNSS in Medium Earth Orbit (MEO), such satellites would reside twenty to forty times closer to Earth, having substantial implications for user performance and satellite payload cost. LEO satellites provide robust accuracy, in part through rapid estimation of carrier phase ambiguities via speedier motion across the sky. Proximity to Earth leads to potentially stronger signals for the end user, giving better tracking performance and substantial resilience in the face of RF interference. Lastly, such a signal is not bound by legacy systems and can be designed with encryption and data authentication for resistance to spoofing attack.

Traditionally the domain of government organizations, building such a satellite navigation service is unprecedented. However, there is also revolution underway in the space sector. The New Space movement challenges the traditional approach, resulting in a ten to one hundredfold reduction in the cost per kilogram to orbit [6]. Satellites and components have become commoditized and roll off assembly lines instead of being individually crafted for each mission. These ingredients create an ecosystem where Mega Constellations of thousands of satellites are being constructed by the likes of OneWeb and SpaceX to meet global demand for broadband. Along with existing GNSS infrastructure, this New Space ecosystem provides the elements needed for a viable commercial LEO navigation service.

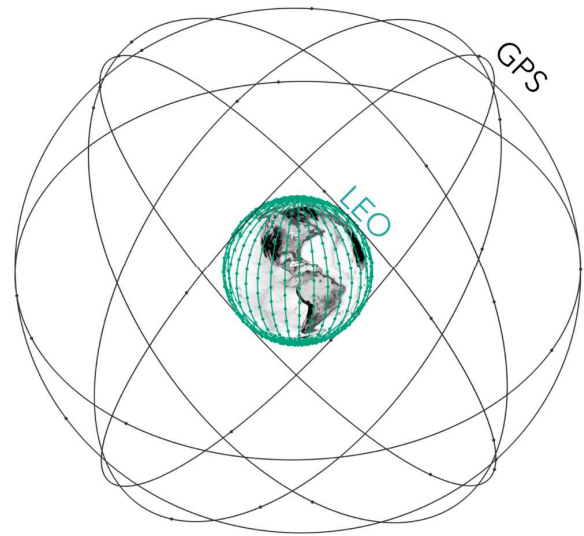

Fig. 1: An example of a 300 satellite Walker LEO constellation compared to the GPS constellation in MEO.

## II. THE DECADE OF THE DECIMETER

To understand the trend in navigation going forward, we first turn to an examination of the past. This section explores historical drivers of navigation needs and the supporting technologies developed to meet new requirements. We further discuss where demands might be headed and the applications driving them today.

Many systems, methods, and tools have been deployed to help humans and machines find their way. New applications pushed the capability of navigation approaches and some demanded new technologies. Fig. 2 shows the progress in Root-Mean-Squared (RMS) positioning accuracy available from contemporary navigation systems throughout the last century. This is not an exhaustive list of every available navigation system, or techniques, but it is representative of the accuracies widely accessible. This shows a clear trend: a tenfold improvement in location accuracy every thirty years.

At the turn of the 20$^{th}$ century, celestial navigation was the state-of-the-art in location technology. Developed in-part to help mariners find their way, it has been used at sea since at least the 16$^{th}$ century [7]. In the early 20$^{th}$ century, this technology was capable of delivering kilometer-level accuracy [8], sufficient for ships to get within sight of land. The Second World War saw rapid growth in aviation and with it the need for new navigation capability in the 1940s. This was met with the emergence of ground-based radio navigation systems such as GEE [9] and later LORAN-C [10] as well as Distance Measuring Equipment (DME) [11]. This supported all weather operation to an accuracy of a few hundred meters. The Cold War brought new accuracy requirements with ballistic missile submarines that could remain submerged for weeks, even months at a time. This required unprecedented accuracy to initialize inertial systems that would run without update for extended periods. This was met with the first satellite navigation system, the U.S. Transit in 1964. Operated by the U.S. Navy, Transit brought tens of meters of accuracy and refined the science of geodesy for continual improvement over its operational lifetime of nearly thirty years [12], [13].

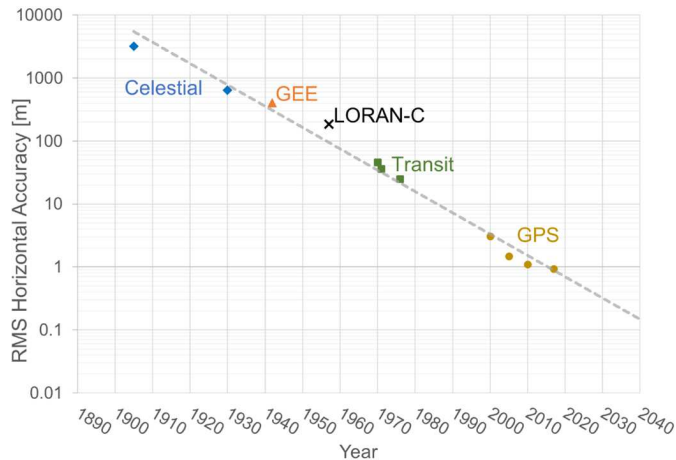

Fig. 2: The progress in location accuracy and associated contemporary technologies over the last century [2]. This shows a clear trend: a ten times improvement in location accuracy every thirty years.

The need for precision continued, where real-time meter-level positioning was desired for fast moving military platforms and precision tactical strikes. One of the mottos of the Joint Program Office of the U.S. Global Positioning System (GPS) was to "drop five bombs in the same hole" [14]. GPS was at full operational capability in 1995 with 24 satellites in MEO. However, it was not until the year 2000 that civilians learned of the system's full potential. Initial degradation of civil signals resulted in position errors that could be as large as 100 meters [15]. Known as Selective Availability (SA), it was in place to prevent enemies from using the system to its full potential. On May 2, 2000, SA was turned off by presidential mandate as the potential economic benefit outweighed the once perceived threat [16]. This enabled meter-level positioning as an open, public, and free service worldwide. The first GPS-enabled phones were released in 2000 and the first GPS smartphone in 2005 [17]. The demand for smartphones and the economies of scale contributed to the now more than 6 billion GNSS-enabled devices worldwide [1].

With every new order of magnitude in position accuracy, a new investment in infrastructure was required. Ground-based radio beacons were erected for aviation, satellites launched for submarines, and now GNSS touches nearly all aspects of modern life. Following the trend in navigation shown in Fig. 2, the mid-2020s are poised for new infrastructure to support decimeter location. Where historically military capability drove innovation and investment in infrastructure for positioning, commercial needs are driving demand today. Decimeter, and even centimeter localization is being sought by autonomous transportation systems such as self-driving cars. Current assessments indicate that autonomous highway driving will require 10 cm RMS accuracy, where city driving will require 5 cm [2]. A representation of the protection levels needed for in-lane positioning is shown in Fig. 3. A diversity of infrastructure is being investigated to meet these needs, from continent-wide GNSS correction services to LiDAR-based localization maps of all major roads. Section III will discuss these developments in more detail.

Along with the advent of the smartphone, GNSS has brought navigation to the masses. Ubiquitous meter-level location is enough for humans to find their way and hence its global impact. Decimeter location is next in this evolution, and it is driven, in-part, by Intelligent Transportation Systems (ITS). Decimeters are needed for humans and robotic systems to coexist and to share the same physical spaces. This will require not only accuracy, but also new capability in terms of guarantees on safety and security.

### III. Localization In Autonomy

Localization is a foundational element of autonomous driving. Knowledge of precise vehicle location, coupled with highly detailed maps, add the context needed to drive with confidence. To maintain the vehicle within its lane, highway operation requires knowledge of location at 50 cm where local city roads require 30 cm [2]. The challenge facing auto makers is meeting the required level of reliability at 99.999999% [2]. This represents one failure per billion miles driven, argued to be representative of Automotive Safety Integrity Level (ASIL) D, the strictest in automotive [2]. This has not yet been demonstrated for road vehicles.

Today, self-driving architectures fall predominantly into two categories: (1) SAE Level 2 driver assistance and (2) SAE Level 4 full autonomy [3]. SAE Level 2 systems are available to consumers today and combine camera-based lane-line detection with radar-based adaptive cruise control for driver assistance in highway environments. In comparison, SAE Level 4 systems are still in the development phase and strive for fully autonomy with no driver input.

Current SAE Level 4 systems primarily rely on LiDAR for localization but also incorporate cameras and radar for both perception and localization [3]. GNSS is not the primary localization sensor due in part to historical availability challenges dating back to the Defense Advanced Research Projects Agency (DARPA) Grand Challenge in 2004 [3], [18]. Some LiDAR localization approaches leverage the surface reflectivity [19]–[21] and others, such as Iterative Closest Point (ICP) [22], the entire 3D structure. Many utilize both for robustness.

High Definition (HD) maps in this context contain a localization layer in addition to layers containing the semantic road information such as the location of lane lines and traffic signals. This localization layer consists of the a-priori surface reflectivity in addition to a possible 3D occupancy map or LiDAR point cloud of the intended driving environment. This results in maps that are substantially more data intensive, with the bulk of the data existing in the localization layer.

In nominal circumstances, LiDAR-based approaches deliver the performance required for automated driving. For example, Liu et al. demonstrated a LiDAR localization system on 1000 km of road data in 2019 [23]. This yielded better than 10 cm, 95% lateral and longitudinal positioning, meeting accuracy requirements for autonomy.

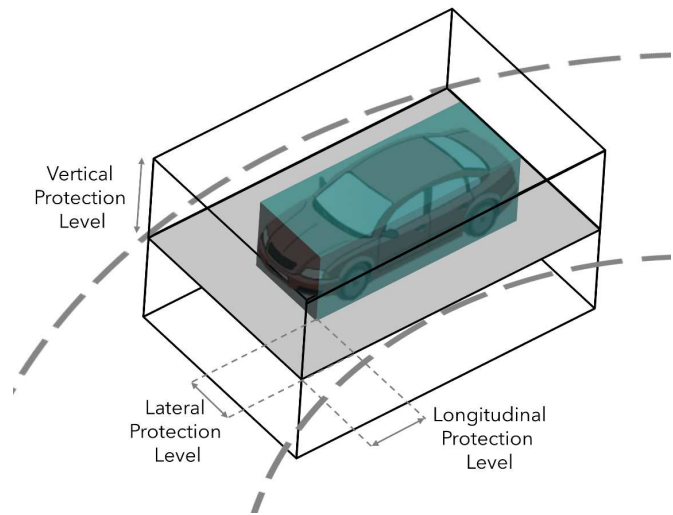

Fig. 3: The lateral, longitudinal, and vertical protection levels required for in-lane positioning and full autonomous driving. Self-driving cars require 20 cm, 95% positioning for highway road geometries and 10 cm, 95% for local streets [2].

Although LiDAR-based localization provides accuracy and availability, LiDAR is not immune to failure. Both LiDAR and computer vision are adversely affected by inclement weather [24]–[28]. Fog, rain, and snow can result in a 25% reduction of LiDAR detection range [26]. Weather conditions further result in a reduction of the number of points per object due to absorption and diffusion [28]. These factors and others have led to the U.S. Department of Transportation stating concerns about LiDAR's ability to function with road snow cover [24]. This concern is compounded by the fact that the U.S. Federal Highway Administration estimates 70% of U.S. roads to be in snowy regions [29].

The 3D structure of the environment can also change with the seasons or with construction, necessitating frequent updates to the LiDAR localization map [30]. Sparse environments with limited distinguishing structure, like open highways, can also lead to poor LiDAR localization performance [30]. Furthermore, like all sensors, LiDAR can suffer from occlusions, for example, by large surrounding vehicles, which block access to the a-priori information contained in the map.

To mitigate the shortcomings of LiDAR as the primary sensor for localization, it is augmented with computer vision, inertial measurement, and odometry inputs [3]. A variety of computer vision approaches to localization have been proposed [31]. Some methods rely on semantic maps [32], global-feature maps [33], landmarks [34], and 3D LiDAR maps [35], [36], while others can operate without a map at all [37]. Odometry inputs are derived from sources including radar Doppler, visual odometry, LiDAR odometry, and wheel speed encoders.

Precision GNSS is complementary to LiDAR. GNSS' microwave signals are unaffected by rain, snow, and fog. GNSS also performs best in open sparse environments like highways. Because of this synergy, Baidu's Apollo framework utilizes a LiDAR + IMU + GNSS localization solution [30]. In test drives with the Baidu system, LiDAR-only localization reaches the alert limits required for autonomous city driving only 95% of the time. The inclusion of an IMU boosts this to 99.99% and with precision GNSS to 100% within the available test drive data. This is substantial since the joint approach strives to address the long tail of localization errors.

GNSS technologies have also seen substantial investment in the Advanced Driver Assistance System (ADAS) and autonomous driving domains. There is now continent-scale deployment of GNSS monitoring stations in service, delivering correction services via cellular. The accuracy required for lane-determination to support future ADAS applications is nearing production where research systems are approaching that needed for full self-driving [3].

Though GNSS accuracy is approaching the needs of autonomy, other risks remain with interference and cybersecurity. RF interference is a growing threat on the road. Often motivated by privacy concerns, low-cost GNSS jammers are a popular means of disrupting fleet tracking. Between 2016 and 2018, more than 50,000 such disruptions were recorded in Europe alone [4], [38].

The cybersecurity of GNSS is another emerging concern. Civil GNSS signals are unencrypted and unauthenticated, leaving vulnerabilities that can be exploited with counterfeit (spoofed) signals. Just less than ten years ago, GNSS spoofing required specialized expertise and equipment costs of upwards of $50,000 [39]. Now, with open source software and more accessible hardware, spoofing attacks can be accomplished for as little as $100 [39].

With increasing accessibility to spoofing has come higher frequency and severity of attacks. For example, at the 2019 Geneva Motor Show, several automotive manufactures including Audi, Peugeot, Renault, Rolls-Royce, Volkswagen, Daimler-Benz, and BMW reported their vehicles' GNSS to be in Buckingham, England in the year 2036, the result from a suspected widespread spoofing attack [40]. Later in 2019, researchers at Regulus Cyber demonstrated this vulnerability in autonomy with a staged spoofing attack on a Tesla Model S and Model 3, creating unsafe behavior of the autopilot [5].

Ultimately, LiDAR, radar, computer vision, inertial sensors, and satellite navigation will require combination to leverage their individual strengths in creating a truly reliable system fit for safety critical autonomous operation. Challenges remain with each of these sensors in achieving the needed availability, integrity, and security in all weather and desired operating conditions. In the next section, we outline an infrastructure approach for robust, precise, and secure satellite navigation to address the needs of autonomy.

## IV. NEW SPACE FOR NAVIGATION

In recent decades, the New Space movement has led to a paradigm shift in aerospace. Driven by commercial needs, non-traditional aerospace ventures have worked to develop faster and cheaper access to space. This model has led to new possibilities, including the Mega Constellations proposed by the likes of OneWeb[1], SpaceX, Telesat, and Amazon for global broadband internet. With new launch providers and tens of thousands of satellites proposed, this has created circumstances for revolution in several industries. Though satellite navigation has traditionally been the arena of governments, the elements for commercial satellite navigation are now present. In this section, we present a concept for a commercial navigation service that meets the needs of autonomous systems under the New Space model and discuss the relevant trade space in the space segment.

### A. New Space

Traditional aerospace is dominated by mission-centric risk adverse programs, resulting in high costs and long development times. The New Space approach is one associated with Silicon Valley which questions the conventional and searches for major strides, rather than baby steps [41]. The New Space philosophy transforms the space market demand into products and services, often accepting higher than traditional levels of risk [42]. The result is new launch providers, new satellite manufacturing techniques, ground stations as a service, and new models for financing space ventures [41].

Perhaps the poster child of this movement is SpaceX, which has challenged the traditional aerospace approach and successfully driven down the cost per kilogram to orbit. Compared to the Space Shuttle, the SpaceX Falcon 9 rocket offers a twentyfold reduction in the cost per kilogram to LEO [6]. Adjusted for inflation, launch costs remained relatively fixed between 1970 and 2000, where many of these launch systems are still in service today [6]. The Falcon Heavy offers an additional twofold reduction in cost, and rocket reusability the potential for a further twofold or more cost reduction [6]. The net result is a near hundredfold reduction in cost per kilogram to LEO.

In addition to disruption in launch services, there has been substantial innovation in small satellite technology and design philosophies. The movement is towards standardization and commoditization of satellite buses, subsystems, and Commercial-Off-The-Shelf (COTS) components. Rather than a focus on specialized space parts, careful selection and trial of industrial and automotive-grade components can yield highly capable satellites at a lower cost and faster development cycle [43]. One such small satellite standard is the CubeSat introduced in 2000 [44], [45], though others have been proposed [46]. Standard CubeSats consist of "Units" that are 10 cm x 10 cm x 11.35 cm, designed to provide 1 liter of useful volume. This has become a widespread industry standard, where 3U, 6U, and even 12U satellite buses are now a commodity offered by the likes of Pumpkin, Blue Canyon, and NanoAvionics [47]. To highlight the capability of these platforms, consider that Planet's more than 200 Earth imaging satellites (Doves) [48] are 3U satellites [49]. This represents nearly ten percent of the 2,218 operational satellites in Earth orbit today [48] and is the New Space paradigm in action.

Low-cost access to space along with commoditized satellite buses and components opens new possibilities, and the potential for revolution. One such movement is underway with the so-called Mega Constellations proposed by the likes of OneWeb, SpaceX, Telesat, and Amazon. Driven by the demand for broadband, tens of thousands of small satellites are planned for LEO. OneWeb, in partnership with Airbus, is producing satellites on assembly lines at a rate of two per day [50], where 74 satellites are already on orbit towards an end goal of up to 900 [51]. SpaceX is leading the charge with 422 Starlink satellites on orbit, where 60 are launched at a time at a possible three-week cadence [52]. Starlink is the most audacious with plans for 42,000 satellites, and already is the largest satellite constellation on orbit [53]. Telesat has partnered with the Canadian government to build a constellation of 300 satellites for initial broadband service in 2022 [54], [55]. Amazon's project Kuiper announced their ambition for 3,236 satellites in April 2019 [56]. Though Amazon has nothing on orbit, it has a head start with Amazon Web Services (AWS) Ground Station, designed to support the traffic of Mega Constellations as a service [57]. Combined, these Mega Constellations represent more than five times the total number of space objects launched since the first satellite, Sputnik 1, in 1957 [58]. Details of these systems are summarized in Table 1.

TABLE 1: MAJOR CONTENDERS FOR BROADBAND LEO MEGA CONSTELLATIONS. BASED ON [53]–[56], [59]–[61].

| Constellation | Num. Sats in Final Design | Num. On Orbit | Altitude [km] | Frequency Band(s) | Planned Initial Service |
|---|---|---|---|---|---|
| OneWeb[1] | 600 – 900 | 74 | 1,200 | $K_a$, $K_u$ | 2021 |
| SpaceX (Starlink) | 800 – 42,000 | 422 | 340 – 1,150 | $K_a$, $K_u$, V | 2020 |
| Telesat | 300 | 1 | 1,000 – 1,200 | $K_a$ | 2022 |
| Amazon (Kuiper) | 3,236 | - | 590 – 610 | $K_a$ | - |

### B. Navigation from LEO

With LEO Mega Constellations entering production for global broadband, it begs the question, what might be possible for satellite navigation under the New Space model? Much work has been done in the domain of LEO-based satellite navigation. In this section, we describe this progress along with the current capability of commercial LEO Position, Navigation, and Time (PNT) services already available.

Compared to GNSS in MEO, LEO offers several distinct advantages and trades as described in [62]. Up to forty times closer to Earth, LEO offers nearly 30 dB (1000x) less zenith path loss as shown in Fig. 4, offering the potential for stronger signals and hence resilience to radio interference. Passing overhead in minutes compared to hours, this rapid geometry change offers observability for rapid convergence of carrier phase differential precise positioning. The trade-off is satellite footprint, it takes nearly tenfold more satellite in LEO to obtain the satellite visibility of GNSS in MEO.

---

[1] At the time of publication, OneWeb announced it had filed for Chapter 11 bankruptcy.

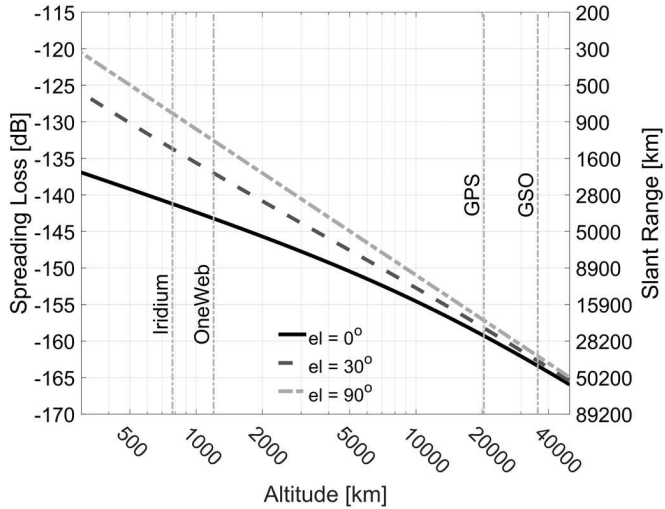

Fig. 4: Free space path loss as a function of orbital altitude.

In the mid-1990s, plans for global cellular 'Big' LEO Constellations such as Iridium [63], Globalstar [64], and Orbcomm [65] were underway, leading to interest in their use for navigation. Rabinowitz et al. examined the benefits of a GPS + LEO system for rapid resolution of integer ambiguities for carrier-phase differential precise positioning [66], [67]. Joerger et al. examined the integration GPS and Iridium for precision and integrity (iGPS) [68], [69].

The Iridium-based Satellite Time and Location (STL) service became operational in May of 2016 [70], [71]. Built by Satelles in partnership with Iridium Communications Inc., many from industry and government are already using this service. This system has demonstrated a positioning accuracy of 20 meters and timekeeping to within 1 microsecond, all in deep attenuation environments indoors, showcasing the improved signal strength from LEO [70].

Morales et al. investigated LEO communications satellite signals of opportunity for navigation including inertial aiding [72]–[74]. There is also commercial interest in LEO communication signals of opportunity. In 2019, Globalstar, in partnership with Echo Ridge, announced the joint development of their Augmented Positioning System (APS) [75] which uses satellite communication signals, not specialized navigation signals, to produce accurate PNT information [76].

The concept of a low-cost navigation payload intended for LEO was first introduced by the authors in [77]. This work showed that the performance of GPS could theoretically be matched with low-cost and space flown COTS components hosted on a LEO Mega Constellation of the type proposed by OneWeb. The Luojia-1A demonstration satellite showcased the possibilities of such a low-cost system, yielding useful navigation signals from LEO in 2018 [78], [79].

Recent work has examined concepts for improved Precise Point Positioning (PPP) through devoted LEO constellations for augmentation of GNSS [80]–[82]. These constellations range in size from 60 to nearly 300 satellites, placed at altitudes between 780 km and 1200 km.

### C. LEO for Autonomous Navigation

Data signed and authenticated with a secure and standardized source of time and location empowers autonomous systems through reliable data sharing and in building collective situational awareness for safe and collaborative autonomy at scale. Along with correction services, GNSS and other technologies are on track to deliver the accuracy needed for autonomy. However, as discussed in Section III, the major missing elements in the combination of GNSS, LiDAR, radar, and vision are guarantees on integrity and cybersecurity.

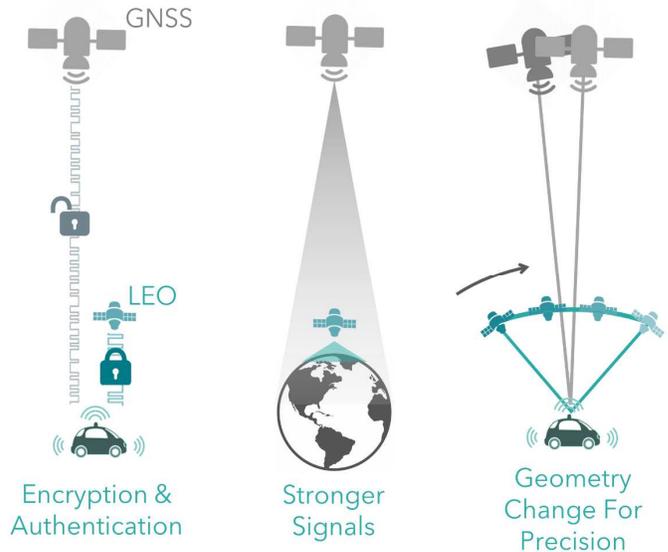

Fig. 5: Desired properties of a LEO navigation service for intelligent transportation systems: encrypted signals with data authentication for resistance to spoofing attack, stronger signals for resilience against interference, and precise positioning aided by rapid geometry change.

A LEO-based navigation service holds potential to provide the backbone to meet the navigation demands of Intelligent Transportation Systems (ITS). When combined, the elements discussed in the review of LEO navigation give rise to precision through rapid carrier-phase ambiguity resolution, robustness to interference through stronger signals, and higher availability through more satellites. Unbounded by legacy signals, a LEO service can further introduction of novel signal encryption and authentication for resistance to spoofing. The desired properties of such a LEO navigation service are summarized in Fig. 5.

With a potential blank slate, what might be desirable from new navigation signals beyond the inclusion of encryption and authentication? It has been suggested by van Diggelen [83] that future navigation services might benefit from signal simplification. Today, each GNSS system has at least a half-dozen signal components, many targeting the needs of specific users. Since each signal consumes power at the satellite, reducing this to two signals could yield higher efficiency and either a smaller and simpler satellite or more powerful signals.

What might such a satellite constellation look like? To estimate satellite size and numbers, we will begin with first principles. Satellite numbers are derived based on the visibility and geometry they yield to terrestrial users. Fig. 6 shows the global Position Dilution of Precision (PDOP) that would result from LEO Polar Walker constellations ranging in altitude from 600 km to 1400 km. To match the performance of GPS, approximately 300 satellites would be required if an altitude between 800 and 1000 km were chosen. It should be noted that the optimal constellation for LEO navigation may not be a Polar Walker configuration, like the Mega Constellations, several other forms may be considered [54].

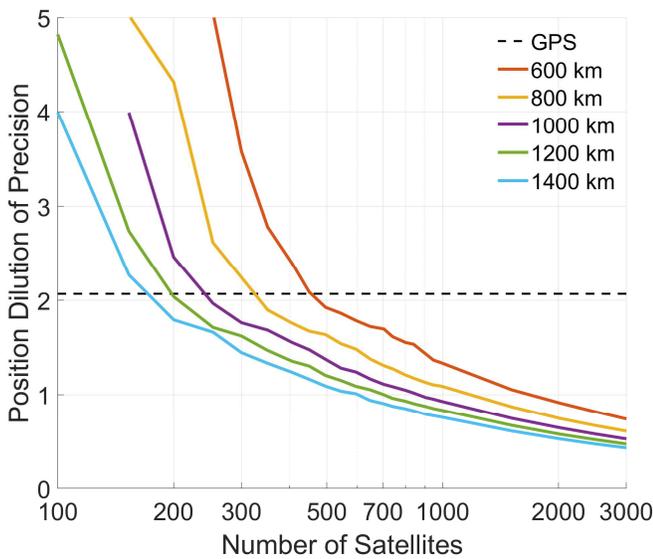

Fig. 6: Global Position Dilution of Precision (95th percentile) as a function of Polar Walker Constellation size and altitude. This assumes a 5° elevation mask.

Satellite size is a function of the required payload power. Consider that Galileo satellites host a 900 W navigation payload [84]. In part, this consists of two rubidium (35 W / unit) [85] and two hydrogen maser clocks (70 W / unit) [86], which combined represent 200 W. This also generates as many as 10 signal components including E1-A, E1-B, E1-C, E5a-I, E5a-Q, E5b-I, E5b-Q, E6-A, E6-B, and E6-C [87] where the combined output power of the Galileo FOC satellites ranges between 254 – 273 W [88]. With a power amplifier efficiency of 51% [89], we can estimate the load on the system as being approximately 535 W for navigation signals alone. On average, per signal, this represents approximately 53 W of power draw on the bus.

To simplify a navigation payload intended for LEO, we begin with removing the atomic clocks. As suggested by [77], GNSS in MEO can be leveraged in LEO to act as the satellite long-term frequency standard. Other clocks such as Oven Control Crystal Oscillators (OCXO) or Chip-Scale Atomic Clocks (CSAC) can be considered for shorter term holdover. This concept is shown in Fig. 7. A GNSS receiver onboard the satellite has further implications for orbit determination, where sub-decimeter GNSS-based orbit determination has been demonstrated in real-time [90].

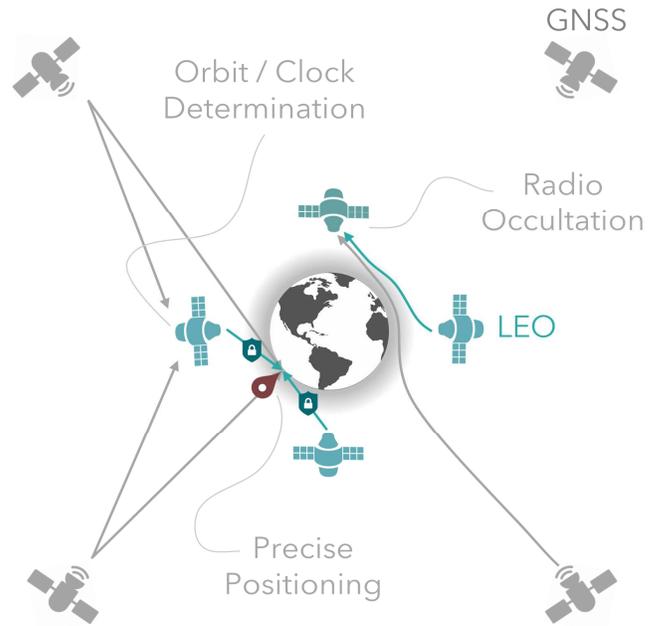

Fig. 7: Possible architecture of a LEO navigation service.

Following [83], simplifying to two signal components can further reduce the required payload power. Since on average, each signal requires approximately 53 W to produce, with two signals plus overhead, this could yield a 100 – 200 W navigation payload for LEO. To examine implications for the end user, first principles has us look at Watts per square meter. LEO satellite footprints are smaller and hence spread this energy over a smaller area, resulting in a net gain compared to MEO. This effect is shown in Fig. 8. For a signal that is broadcast to the horizon, this gain is 6 – 10 dB (4 – 10x). For a spot beam that users can only see at elevations greater than 30 degrees, this can be 13 – 20 dB (20 – 100x). Assuming a broadcast to the horizon, it appears that a 100 – 200 W navigation payload could yield signals 4 – 10 times stronger to end users, assuming an even distribution of power. Scaling also indicates that GNSS-level signals can be achieved with 4 – 10 times less payload power, or approximately 20 – 50 W with overhead. To put satellite size in context at these power levels, the OneWeb 150 kg Arrow satellite bus can support a 200 W payload [91] while some 12 kg 6U CubeSats can support nearly 30 W [92].

How much more power is useful for ITS? The Iridium-based STL delivers +30 dB (1000x) stronger signals, giving signal access to stationary indoor users [70]. For dynamic vehicles requiring precise positioning, indoor navigation might be better served by other technologies. In the author's personal experience with autonomous vehicle on-road testing, two culprits caused the most GNSS disruptions: the tree canopy and RF interference. Table 2 shows the gains in these categories as a function of additional signal power. For simplicity, this table assumes L-band for a direct comparison to GNSS. In terms of material penetration, +5 dB is enough for deciduous trees, +10 dB for wooden walls and redwoods, +20 dB for most tree canopies and most walls, and +30 dB for multiple walls. It should be noted, however, that even with penetration, challenges will remain with multipath. In terms of jamming mitigation, consider the effective radius of a high output civil GNSS jammer, the type typically used to maintain privacy from fleet trackers on the road. At 500 mW, the effective radius drops from nearly a kilometer to less than the length of a city block at +20 dB and less than four car lengths at +30 dB.

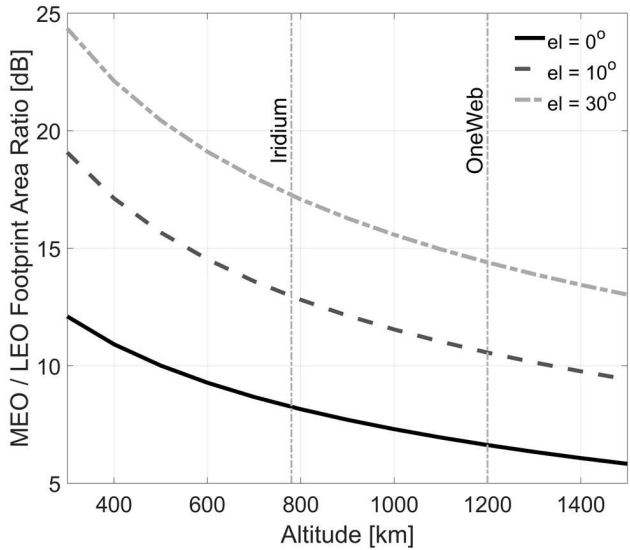

Fig. 8: Satellite footprint area ratio of MEO (Galileo) and satellites at LEO altitudes. To first principles, this shows the net gains of satellites in LEO in terms of signal transmit power. The different elevation masks represent satellite beamwitdths, where narrower beams spread energy over a smaller area.

What precision is needed? Fully autonomous driving requires 10 cm, 95% positioning [2]. This will be required instantly; convergence times longer than a few seconds will be unacceptable. PPP-RTK GNSS corrections are already deployed at continent-scales, offering rapid convergence with regional atmospheric corrections [3]. Compared to GNSS in MEO, LEO offers rapid geometry change, already accelerating the convergence of PPP from tens of minutes to less than 1 minute while unaided by atmospheric correction information. This still is not fast enough, indicating that some form of atmospheric corrections will be needed for a LEO-based system. These could be derived in the same way that corrections are today, with networks of ground monitoring stations. However, there is also the growing commercial weather data market which could yield a mutually beneficial relationship with PNT.

Once the domain of governments with missions such as COSMIC [93], GNSS Radio Occultation (RO) is a critical input to global weather models. This is transitioning to the commercial sector, where NOAA and other government agencies are moving towards purchase of RO and other data. As commercial constellations grow from 'Big' to 'Mega' this gives the opportunity not only for more GNSS to LEO-RO soundings but for LEO to LEO RO soundings as well. Multi-frequency gives ionospheric measurements and RO bending angles a measure of water content in the troposphere. These are ingredients for atmospheric corrections for precise positioning. This can create a positive feedback loop of improved navigation signals for atmospheric modeling and improved atmospheric modeling for enhanced positioning.

To create such a LEO navigation service would be cost prohibitive if additional LEO signals came with the price tag of current GNSS satellites in MEO. The latest GPS III satellites are $345M each [94]. In comparison, OneWeb Arrow satellites are around $1M [95]. By leveraging the GNSS constellation in MEO for clock and orbit determination, a simplified navigation payload for LEO could be constructed to yield stronger signals with encryption and authentication. On a constellation of 300 satellites, this would add the satellite visibility and geometry equivalent to one of GPS, Galileo, GLONASS, or BeiDou. Moreover, in the New Space ecosystem, such a constellation could conceivably be built for less than the cost of a single GPS satellite.

TABLE 2: SATELLITE NAVIGATION SIGNAL POWER, MATERIAL PENETRATION, AND JAMMING MITIGATION. TREE CANOPY PENETRATION INFORMATION IS BASED ON [96]. MATERIAL PENETRATION INFORMATION IS BASED ON [97].

| $C/N_0$ Margin Over GPS [dB-Hz] | Material Penetration (L-Band) | | | | | | Jamming Mitigation | |
|---|---|---|---|---|---|---|---|---|
| | Tree Canopy | # Walls Wood (-10 dB) | # Walls Brick (-12 dB) | # Walls Reinforced Concrete (-15 dB) | Heat Protected Glass (-17 dB) | Shipping Container (-25dB) | 500 mW* Jammer Effective Radius [m] | Jammer Needed for 100 m** Radius |
| 0 | Limited | 0 | 0 | 0 | 0 | 0 | 750 | 10 mW |
| 5 | Deciduous | 0 | 0 | 0 | 0 | 0 | 430 | 32 mW |
| 10 | Redwoods | 1 | 0 | 0 | 0 | 0 | 240 | 100 mW |
| 20 | Most | 2 | 1 | 1 | 1 | 0 | 80 | 1000 mW |
| 30 | Most | 3 | 2 | 2 | 1 | 1 | 20 | 10,000 mW |

* 500 mW is on the upper end of common GNSS jammers found on the road today known as a Personal Privacy Device [98].
** 100 m is approximately a city block, but shadowing from buildings, etc, makes a city block more complicated as a metric.

## V. Conclusion

Following historical trends of new infrastructure for new navigation capability, the mid 2020s forecast new investment to deliver widespread decimeter, or better, positioning. Intelligent Transportation Systems including self-driving cars and autonomous aerial systems require this precision to function. Furthermore, their effective interoperation necessitates a common standard. The safest maneuvers are informed with the most complete picture of the surroundings. This requires going beyond the line of sight of vehicle sensors and creating situational awareness at municipal levels and beyond. This demands an environment of collaborative data sharing through vehicle communication and broadband connectivity. This allows vehicles and infrastructure to act collectively, improving safety and reducing the risk of collision. Such data sharing is only effective if there is an agreed upon common standard and datum with appropriate measures for data security.

To meet these demands, we present a concept for new navigation infrastructure in Low Earth Orbiting (LEO). The navigation benefits from LEO are many. Compared to GNSS in MEO, LEO satellites reside twenty to forty times closer to Earth, having substantial implications for user performance. LEO satellites can provide robust accuracy, in part through fast convergence of precise positioning resulting from speedier motion across the sky. Proximity to Earth further leads to stronger signals for the end user, giving better tracking performance and resilience to radio interference. Such a signal is not bound by legacy and can be designed with encryption and data authentication for enhanced security and resistance to spoofing attack. Data signed and authenticated with a secure and standardized source of time and location empowers autonomous systems through reliable data sharing in building collective situational awareness for safe autonomy at scale.

A LEO constellation of approximately 300 hundred satellites can provide similar coverage to GPS today. With broadband Mega Constellations of tens of thousands of satellites being constructed by the likes of OneWeb, SpaceX, Telesat, and Amazon, highly capable satellites and components are now made in volume on assembly lines, creating an ecosystem for previously unimagined space infrastructure. Under this revolution, a LEO navigation constellation can conceivably be built for the cost of a single GPS III satellite.